\documentclass[conference]{IEEEtran}
\IEEEoverridecommandlockouts
\usepackage{cite}
\usepackage{amsmath,amssymb,amsfonts}
\usepackage{algorithmic}
\usepackage{graphicx}
\usepackage{textcomp}
\usepackage{xcolor}
\usepackage{svg}
\usepackage{mathrsfs}
\usepackage{calligra}
\usepackage{float} 
\usepackage{booktabs}
\usepackage{subfig} 
\usepackage{pgfplotstable}
\pgfplotsset{compat=1.7}
\pgfplotsset{minor grid style={line width=.1pt,draw=gray!20}, major grid style={line width=.2pt,draw=gray!50}}
\usepackage{tikz, pgfplots}
\def\BibTeX{{\rm B\kern-.05em{\sc i\kern-.025em b}\kern-.08em
    T\kern-.1667em\lower.7ex\hbox{E}\kern-.125emX}}

\begin{document}

    \title{Application of the List Viterbi Algorithm for Satellite-based AIS Detection  \\
        {\footnotesize \textsuperscript{}}
    }

    \author{ Linda Kanaan\textsuperscript{1,2}, Karine Amis\textsuperscript{1}, Frédéric Guilloud\textsuperscript{1}, Rémi Chauvat\textsuperscript{3} \\

    $^1$ IMT Atlantique, Lab-STICC, 29238 Brest, France,\\
    $^2$ TéSA, 31000 Toulouse, France,\\
    $^3$ Kinéis, 31520 Ramonville Saint-Agne, France \\

    emails: \{linda.kanaan, karine.amis, frederic.guilloud\}@imt-atlantique.fr, rchauvat@kineis.com
     \thanks{This work has been financially supported by Kinéis, CNES and Région Bretagne.}}
    
    \maketitle
    
    \begin{abstract}
    Satellites receiving Automatic Identification System (AIS) packets in dense areas are particularly prone to AIS channel overload due to the extensive number of vessels. Thus a failure of detection might be caused by the collisions among AIS messages. To improve the detection capability, we propose to exploit the presence of the cyclic redundancy check (CRC) in AIS frames by using the parallel list Viterbi algorithm (PLVA) instead of the classical Viterbi algorithm (VA) often used for decoding AIS signals. The performance of combining the PLVA with AIS post processing including the CRC is studied with two detectors, one coherent and the other differential, in two channel models: a single-user AWGN channel and a more realistic multiple-access AIS channel. We also show the impact of the PLVA parameters on the success recovery rate. The simulation results show that the resulting procedure can significantly improve the packet error rate (PER) at the cost of a limited increase of the computational complexity. The proposed technique could be applied to improve the performance of interference cancellation receivers by significantly lowering the AIS decoding threshold.
    \end{abstract}

    
    \section{Introduction}

        AIS \cite{Technical} is employed in the maritime sector to ensure vessel traffic safety. The AIS signal is generated and broadcast via standardized very high frequency (VHF) transceivers, thus allowing ships to automatically exchange information with other ships and nearby coastal stations up to 20 to 30 nautical miles. 
        
        Starting from 2007, AIS receivers have been embedded into low earth orbit (LEO) satellites (Sat-AIS) \cite{SATais}, enabling global supervision of the maritime traffic. Since AIS was mainly developed for terrestrial communications, the Sat-AIS reception has been accompanied by several challenges including message collisions, propagation delay, attenuation and Doppler effect. 
        In environments with high vessel density, the likelihood of AIS message collisions increases because vessels are unable to coordinate their transmissions beyond a distance way below the footprint of the satellite, leading to a higher probability of undetected or incorrectly decoded signals. This necessitates the implementation of highly reliable decoding mechanisms to ensure the integrity of the transmitted information. 
        
        Several approaches have been considered for mitigating the interference for AIS detection, this includes antenna array processing \cite{phasedantenna1,phasedantenna2}, bandwidth separation in sub-zones \cite{threezones2} and interference cancellation approaches \cite{jointMLSE,PIC1,SIC2}. It is important to note here that these three approaches could be combined altogether. However, all these approaches are intrinsically limited by the poor energy efficiency of the (uncoded) AIS signals, forbidding correct decoding except for large values of received signal-to-noise plus interference ratio. Whereas previous works \cite{ap1},\cite{ap2} exploit a priori information on AIS message bits based upon statistical analysis and knowledge of satellite position, we assume uniformly distributed bits in this paper. However, exploiting this knowledge could be combined for further improvement.

        
        
        In this paper, we intend to improve AIS decoding by generating candidates diversity (i.e. sequence diversity) with a later perspective of improving the performance of receivers including interference cancellation. The AIS frame definition constraints including the CRC will be used to ensure the validity of each candidate. Prior works have already proposed to exploit the CRC for error correction in AIS. In \cite{threezones2}, two simple yet effective strategies are proposed, bit flipping and exploiting syndromes of CRC codewords. Both do not use the trellis structure, instead, soft demodulated outputs are exploited. In \cite{jointMLSE}, maximum likelihood sequence estimation (MLSE) including CRC and bit stuffing constraints is performed. This yields the best performance in the single user approach so far, but it has significant computational complexity. To the best of our knowledge, the application of the list Viterbi algorithm (LVA) to the AIS context has not been analyzed so far.  Due to the capability of LVA to maintain multiple candidate paths, researchers have been able to achieve significant advancements in accuracy and reliability in different fields. Different applications of LVA along with an outer code for error detection (typically CRC) appeared in the literature \cite{listviterbi,listviterbi2,listvit5,listvit6}. Parallel LVA (PLVA) has been proposed in \cite{listviterbi,listviterbi2} and improved in \cite{listvit7}. An alternative to the PLVA is the serial LVA (SLVA) proposed in \cite{listviterbi,listvit3,listvit4}.

        As previously mentioned, our objective is to reduce the packet error rate (PER) thanks to candidates diversity. To achieve this, PLVA is implemented instead of the VA \cite{viterbi} for the Gaussian minimum shift keying (GMSK) demodulation. The contributions of this paper are (i) Application of PLVA for AIS signals detection in a single user AWGN channel (ii) Optimization of the  PLVA parameters for the optimized coherent and differential detection of AIS signals (iii) Simulations with optimized PLVA for realistic AIS signals parameters.
        The remainder of this paper is organized as follows. Section II gives an overview of the AIS medium access scheme and the consequences of satellite reception. It also describes the AIS packet structure and system model. Section III presents the decoders implemented for the detection of AIS signals. This is followed by the LVA principle combined with the AIS message post-processing in Section IV. Simulation results are provided in Section V. Finally, a conclusion is drawn in Section~VI.
    
    \section{System Model} 
    
        \subsection{AIS access protocol}
        \label{AIS collision}

            AIS is a synchronized network using time slotted access to its radio channels. Self-organized time division multiple access (SOTDMA) constitutes AIS main access scheme. Under this mode of operations, vessels are avoiding the selection of the same time slots by listening to the AIS channels and advertising in their messages the next slots they intend to use. 
        
            When the vessels are sufficiently close to each other, there is likely no collision among the transmitted signals due to the SOTDMA scheme 
            and the differential propagation delays much inferior to the AIS buffer length located at the end of the AIS packet (see Fig.~\ref{fig: aispacket_updated(1)}~). However, if the AIS signals are transmitted from vessels that are far from each other within the satellite footprint, two types of collision could occur.
            
            The signal collision due to uncoordinated transmissions targeting the same time slot in two different SOTDMA cells leads to a full collision among the signals. The interference resulting from this type of collision is denoted here "Class 1 Interference". The number of ships within satellite detection range and the rate at which ships report determine how many collisions of this kind occur during each time slot. In the case of a satellite reception with large ground footprint, the number of interfering signals can potentially be very important especially in dense maritime areas.
            
            Furthermore, even if the network operates in a synchronous manner (i.e. with temporal slots synchronized on UTC time), the different propagation delay within signals can generate partial collisions between adjacent slots. The interference resulting from this type of collision is denoted here "Class 2 Interference". It has less impact on the decoding performance.

        \subsection{AIS packet}
        
\begin{figure}[t]
                \centering
                 \includegraphics[width=\columnwidth]{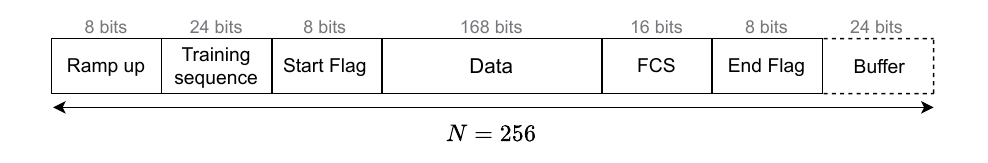}   
                \caption{AIS packet}
                \label{fig: aispacket_updated(1)}
            \end{figure}

            The ITU-R M.1371-5 standard \cite{Technical} covers the description of the AIS signal format transmitted by the vessels. Briefly, an AIS packet is $N$ bits long where $224 \leq N\leq 256$. An AIS slot has a duration of 256 AIS symbols and assuming perfect SOTDMA, a single AIS packet should be transmitted per slot. An AIS packet contains signaling and data bits following the structure of Fig. \ref{fig: aispacket_updated(1)}.

            After the transmission of the AIS data (168 bits), the cyclic redundancy check (CRC) is computed and appended in the FCS field, thus creating the data message part of the AIS packet. Bit stuffing is then performed in order to avoid long ’1’ sequences and the existence of the ’01111110’ pattern reserved for the start flag and the end flag. Afterward the message is concatenated to the signaling parts of the packet including the training sequence with the start and end flags of the message. Next, it is encoded through non-return to zero inverted (NRZI) and modulated with GMSK as described hereinafter.
            
        \subsection{Transmission Signal Model}
        
            AIS signals transmit data using the GMSK modulation \cite{Digitalpm}, a particular form of continuous phase modulation (CPM) that we describe in details in this section. We consider a sequence of $\textit{N}$ independent and identically distributed (i.i.d.) symbols $\mathbf{a}=\{a_i \}_{0 \leq i \leq \textit{N}-1}$ to be transmitted such that $a_i~\in \mathcal{M}=~\{\pm 1\}$. Let $s(t,\mathbf{a})$ denote the complex envelope of the AIS signal given by
            \begin{equation}
            s(t,\mathbf{a})= \sqrt{\frac{2E_s}{T_s}}e^{j \theta(t,\mathbf{a})}
            \label{eq:1}
            \end{equation}
            where $E_s$ is the average symbol energy, $T_s$ is the symbol duration. $\theta(t,\mathbf{a})$ is the signal phase that depends on the symbol sequence $\mathbf{a}$, defined by:
            \begin{equation}
            \theta(t,\mathbf{a})= 2 \pi h \sum_{i=0}^{N-1} a_i q(t-iT_s)
            \label{eq:2}
            \end{equation}
            with $h$ the modulation index and $q(t)$ the phase shaping pulse.
            In practice, the infinite GMSK frequency pulse support is truncated to $[0, LT_s]$ and referred to as $L$-GMSK where $\textit{L}$ is the CPM memory.     
        
        \subsection{Received Signal Model}
        
            In this paper, it is assumed that the satellite is equipped with just one antenna for reception, and that each vessel has one antenna for transmitting the AIS signals. A slowly varying channel during each AIS signal transmission with pure line of sight (LOS) path is considered. Furthermore, perfect acquisition of signals is assumed and we denote by "detection" the process of estimating data symbols in received AIS signals. The multi-user received signal model during a given AIS time slot can be written in a complex baseband representation as
                        \begin{sloppypar}
            \begin{equation}
            \begin{aligned}
            r(t)=\sum_{i=1}^{\mathscr{N}}\alpha_i& s_i(t-\tau_i,\mathbf{a}) e^{j(2  \pi f_{d,i}(t)t+ \phi_i)} + n(t)\\
            \end{aligned}                          
            \label{eq:4}
            \end{equation}
where $\mathscr{N}$ is the number of colliding AIS signals, $s_i(t,\mathbf{a})$ is the $i^{th}$ AIS signal with $1 \leq i\leq \mathscr{N}$, $\alpha_i$ is the attenuation coefficient of the LOS path, $\tau_i$ is the propagation delay, $\phi_i$ is the phase shift of the $i^{th}$ path and is uniformly distributed in $[0, 2\pi )$. $f_{d,i}(t)$ is defined as $f_{d,i}(t)= f_{d,i}+ \eta_it/2$, it represents the time varying Doppler frequency shift associated with the $i^{th}$ signal with $f_{d,i}$ as the initial Doppler shift and $\eta_i$ standing for the Doppler rate of the signal. $n(t)$ is a complex circular stationary AWGN noise with single-sided power spectral density $2N_0$.

            We assume that for a given user of interest, the receiver has a perfect knowledge of these propagation parameters with perfect carrier frequency and timing synchronization. We neglect the effect of Doppler rate ($\eta_i =0$) having limited impact due to the sufficiently slow frequency variation compared to the AIS message duration, in this case $f_{d,i}(t)= f_{d,i}$.
            \end{sloppypar}

     \section{Detection for Space-based AIS}    
        \subsection{Maximum Likelihood Sequence Estimation Principle \label{sec:mlse}}
        
            The decoding algorithm is described for an AWGN channel model. It will be applied in the multiple access system model when the receiver is perfectly synchronized with the signal of interest and the multiple access interference is considered as an additional noise. In the single-user AWGN channel model, the described coherent detection is optimal. The equivalent baseband received signal is given by:
            \begin{equation}
            r(t) = s(t-\tau,\mathbf{a})e^{j(2 \pi f_{d}t+ \phi)} + n(t)
            \label{eq:7}
            \end{equation}
            with $f_{d}$, $\tau$, $\phi$ and $n(t)$ having the same definition as in (\ref{eq:4}).\\
            The MLSE algorithm maximizes the correlation between $r(t)$ and $s(t ,\Tilde{ \mathbf{a}})$ over all possible realizations for $\Tilde{ \mathbf{a}}$.
            
            \begin{equation}
            \hat{ \mathbf{a}}=  \arg \max_ {\Tilde{ \mathbf{a}}}  \Re\left[ \int{ r(t)  s^*(t-\tau ,\Tilde{ \mathbf{a}})e^{-j(2 \pi f_{d}t+ \phi)}dt}\right]
            \label{eq:8}
            \end{equation}
            where $(.)^*$ is the complex conjugate and $\Re$ represents the real part.
    
    \subsection{Coherent MLSE-Detector of CPM}
        \label{coh}
         In coherent detection, channel parameters need to be estimated by the receiver. In practice, it requires a preliminary synchronization step.
         
          To solve the complexity problem involved by the computation of the different correlation metrics, the Viterbi algorithm \cite{viterbi} is applied using the trellis representation of the CPM. 
            
            




        The phase of the baseband CPM signal in (\ref{eq:2}) can be split into two main terms as follows:
        \begin{equation}
        \begin{split}
        \theta(t,\mathbf{a})&=2 \pi h \sum_{i=n-L+1}^{n} a_i q(t-i T_s) + \pi  h \sum_{i=0}^{n-L}a_i\\
                  & = \Theta_n(t) + \theta_n
        \end{split}    
        \label{eq:11}
        \end{equation}
        
        The first term $\Theta_n(t)$ corresponds to the contribution of the current and last $(L-1)$ symbols, $\theta_n$ describes the contribution of the rest of the past symbols. For the optimal coherent detector, the trellis states are represented by the phase evolution and the previous $(L-1)$ symbols given by the state vector: 
        \begin{equation}
        \sigma(n)= [\theta_n,a_{n-L+1},...,a_{n-1}]
        \label{eq:12}
        \end{equation}
        
        Let $h=u/p$ where $u$ and $p$ are co-prime integers. The number of states is equal to $S=pM^{L-1}$ or $S=2pM^{L-1}$ if $u$ is even or odd respectively.
        
       Given the state $\sigma_i$ at instant $n+1$, the Viterbi algorithm computes the cumulative metric for each possible transition towards this state. Then it selects and stores the previous state which yields the maximum one.
       
       \begin{equation}
      \Lambda_{n+1}(\sigma_i) = \max_{\sigma_j \rightarrow \sigma_i} \left(\{\Lambda_{n}(\sigma_j)+\delta_{n}(\sigma_j,\sigma_i)\}\right)
     \end{equation}
     where $\delta_{n}(\sigma_j,\sigma_i)$ denotes the branch metric from state $\sigma_j$ to state $\sigma_i$ with $i$ and $j$ $\in$ $[\![1,S]\!]$ and is defined as 
     
    \begin{equation}\delta_n(\sigma_j,\sigma_i)=  \Re\left[\int_{nT_s}^{(n+1)T_s}{ r(t) s^*(t-\tau ,\Tilde{ \mathbf{a}})e^{-j(2 \pi f_{d}t+ \phi)}dt}\right].\end{equation}
    
     At the end, the MLSE-based decision is taken by selecting the final state with highest cumulative metric. A backtracking is applied to deliver the corresponding estimated symbol sequence $\hat{ \mathbf{a}}$.
    
        
        \subsection{Optimized Differential MLSE-Detector}
        \label{diff}
        
        Using a differential detector \cite{differential} \cite{optimaldiff}, the phase of the incoming signal is tracked based on the phase difference between consecutive symbols. The received signal is multiplied by a time-delayed and conjugate version of itself, thus this process simplifies the detection problem by eliminating the phase shift $\phi$ coming from the transmission. Let $K$ be the delay factor. The differential signal is given by:
        \begin{equation}
        R_K(t) =r(t)r^*(t- KT_s) = S_K(t-\tau,\mathbf{a})e^{j(2 \pi f_{d}KT_s)} + N_K(t)
        \label{eq:13}
        \end{equation}
        with $r(t)$ defined as in (\ref{eq:7}), $N_K(t)$ depending on both the signal and the AWGN noise\cite{optimaldiff}, and $S_K$ defined as:
        \begin{equation}
              \qquad S_K(t,\mathbf{a})= \frac{2E_s}{T_s}  e^{i(\Theta_K(t,\mathbf{a}))},\\
        \label{eq:14}
        \end{equation}
        where $\quad \Theta_K(t,\mathbf{a})= \theta(t,\mathbf{a}) -\theta(t-KT_s,\mathbf{a})$. With $0 \leq \zeta < T_s$, the differential phase $\Theta_K(\zeta + nT_s,\mathbf{a})$ reads: 
             
        \begin{equation}
        \begin{split}
        \Theta_K&(\zeta + nT_s,\mathbf{a}) =\varphi_n(\zeta) +\theta_n  \\
          \text{with} \quad \varphi_n(\zeta)=2 &\pi h \sum_{i=0}^{L-1} (a_{n-i} - a_{n-K-i})  q (\zeta +iT_s)\\
          \text{and} \quad &\theta_n= \pi h \sum_{i=0}^{K-1} a_{n-L-i}
        \end{split}
        \label{eq:15}
        \end{equation}\vspace{0.2cm}
        \begin{sloppypar}
        These two terms $\varphi_n(\zeta)$ and $\theta_n$ are fully determined  by the set of symbols $a_{n-i} $ with $1 \leq i < L+K-1$.  Thus unlike the trellis of the original CPM signal, in this case $\theta_n$ does not need to be stored as a state parameter.  The state is defined as $\Sigma(n)=[a_{n-L-K+1} , ..., a_{n-1}]$ with $M^{K+L-1} $ different possible states independent from the modulation index. The MLSE detection criterion described in Section \ref{sec:mlse} applies on the resulting differential signal replacing $r$ and $s$ by $R_k$, $S_K$ in (\ref{eq:8}).
The differential detector has been proven to be efficient for GMSK and $L= 3$ with the optimized delay value $K=3$ \cite{optimaldiff}.
        \end{sloppypar}

    \section{Parallel List Viterbi Algorithm for AIS Signal Decoding}
     
    To improve the detection of the AIS signals by increasing the diversity of candidates, we propose to use the LVA, taking advantage of the CRC embedded in the AIS message. We describe in this section the principle of "Parallel LVA" \cite{listviterbi},\cite{listviterbi2} and its implementation details.
    
    \subsection{Principle}
    
    PLVA is an extension of the traditional Viterbi algorithm, it ensures the diversity of candidates by maintaining a list of multiple paths with their corresponding metrics instead of only keeping the most likely path. Two main parameters could be considered for PLVA: 
  ${P}$ The number of paths stored for each state at each stage of the trellis, and
     ${C}$ the number of the most likely candidates chosen at the final stage of the trellis. 
   
    PLVA produces an ordered list of the  best estimates of the transmitted information sequence following the trellis search, this requires computing the $P$ best paths for each state at every trellis section.

    \subsection{Algorithm}
    
\begin{figure}[t] 
       \centering
        \includegraphics[width=0.9\columnwidth] {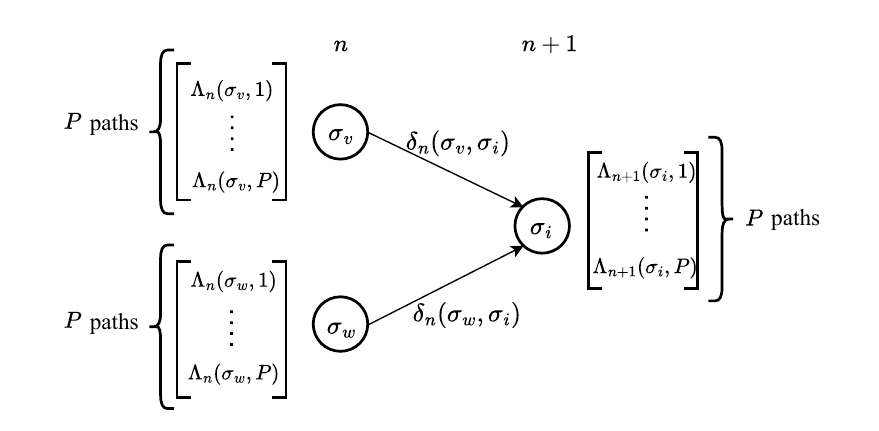}   
        \caption{Cumulative Metrics of the PLVA}
        \label{fig: list viterbi}
    \end{figure}

    For $S$ states, $N$ transmitted symbols and $P$ paths, PLVA requires maintaining a cumulative metric array of the $P$ maximum cumulative metrics and a state array which stores their path history (the previous state and the rank of the previous path) for each time instant in the trellis. 
    For a state $\sigma_i$ at instant $n$, a list sorted in decreasing order of size $P$ is stored. This list contains the $P$ maximum cumulative metrics of the $P$ paths reaching $\sigma_i$ state. This is illustrated in Fig. \ref{fig: list viterbi}. The path with the highest cumulative metric is considered to be the most likely path, the path with the second highest cumulative metric is considered to be the second most likely path and so on.
    
    The cumulative metric of the $k$-th most likely path reaching state $\sigma_i$ at instant $n$ is denoted by $\Lambda_{n}(\sigma_i,k)$. 
    For $n= 0$, decoding starts from a state with only one cumulative metric initialized such that $\Lambda_{0}(\sigma_i,1)= 0$. At instant $n+1$, the length-$P$ cumulative metric vector for state $\sigma_i$ denoted by $\boldsymbol{\Lambda}_{n+1}(\sigma_i) = [\Lambda_{n+1}(\sigma_i, 1), \Lambda_{n+1}(\sigma_i, 2), \hdots, \Lambda_{n+1}(\sigma_i, P)]^T$ where $(.)^T$ stands for transposition, is computed as follows:
     
    \begin{equation}
         \boldsymbol{\Lambda}_{n+1}(\sigma_i) = \max^{(P)}_{\genfrac..{0pt}{2}{\sigma_j \rightarrow \sigma_i}{k \in [\![1,P]\!]}}
          \left(\{\Lambda_{n}(\sigma_j,k)+\delta_{n}(\sigma_j,\sigma_i)\}\right)
        \label{eq:17}
    \end{equation}
    with $\displaystyle \max^{(P)}(\mathcal{S})$ denoting the function that selects the $P$ highest values in the set $\mathcal{S}$.
    
    At the last ($N$-th) trellis section, $C$ candidates with the highest cumulative metrics among the $S  P$  paths stored in the lists of final states are extracted. The PLVA decoding is followed by the post processing by which the NRZI decoding is applied followed by two main criteria for selecting the right candidate. After the stuffing bits retrieval, the length criterion is applied on the decoded frame making sure whether it is equal to 184, which is the length of the AIS data message. If this is not the case, the frame is discarded. Otherwise, if the length condition is satisfied, the selected frame undergoes the CRC check criterion. 
    
    To reduce processing time without affecting performance, a stop criterion can be applied. Candidates are ranked in descending order according to their cumulative metric.  When one of them satisfies all the constraints mentioned above (length-184, valid CRC), it is selected as the decision and the post-processing stops. If none of the candidates meet the two post-processing criteria, an error is declared and no signal is detected.
    
    For small list sizes, the complexity of the PLVA is generally considered acceptable when compared to that of the classical Viterbi Algorithm. Whenever the complexity is critical, several implementations of the LVA are presented in \cite{listvit4} including PLVA, SLVA and its improved versions. With any of these implementations a similar performance could be attained when the list size $P$ is properly tuned. Despite the slightly higher complexity of PLVA, its parallel nature makes it a more straightforward and easier-to-implement option with reduced decoding latency, which is why it is often chosen in practice.

    \section{Numerical Results}
    
In this section, we compare the performance in terms of AIS message recovery of the receiver based on the PLVA with the one based on the VA applied either on the original or the differential received signal. The generic system model is given in (\ref{eq:4}) and we assume perfect channel parameter knowledge at the receiver. We first consider the single user AWGN channel in Section~\ref{Detection in a single-user AWGN channel}. Then we optimize the parameters $(P,C)$ of the PLVA based on a simplified two-user AIS channel. In the last part, we evaluate the performance in a more realistic AIS multiple-access scheme with low-to-high system load. According to AIS signal definition, in the simulation setup we take symbol period $T_s= 1/9600$s, $BT=0.4$, CPM memory $L=3$ and modulation index $h=0.5$. 
    
    \subsection{Detection in a single-user AWGN channel}
    \label{Detection in a single-user AWGN channel}
\begin{figure}[t] 
       \centering
        \includegraphics[ width=\columnwidth]{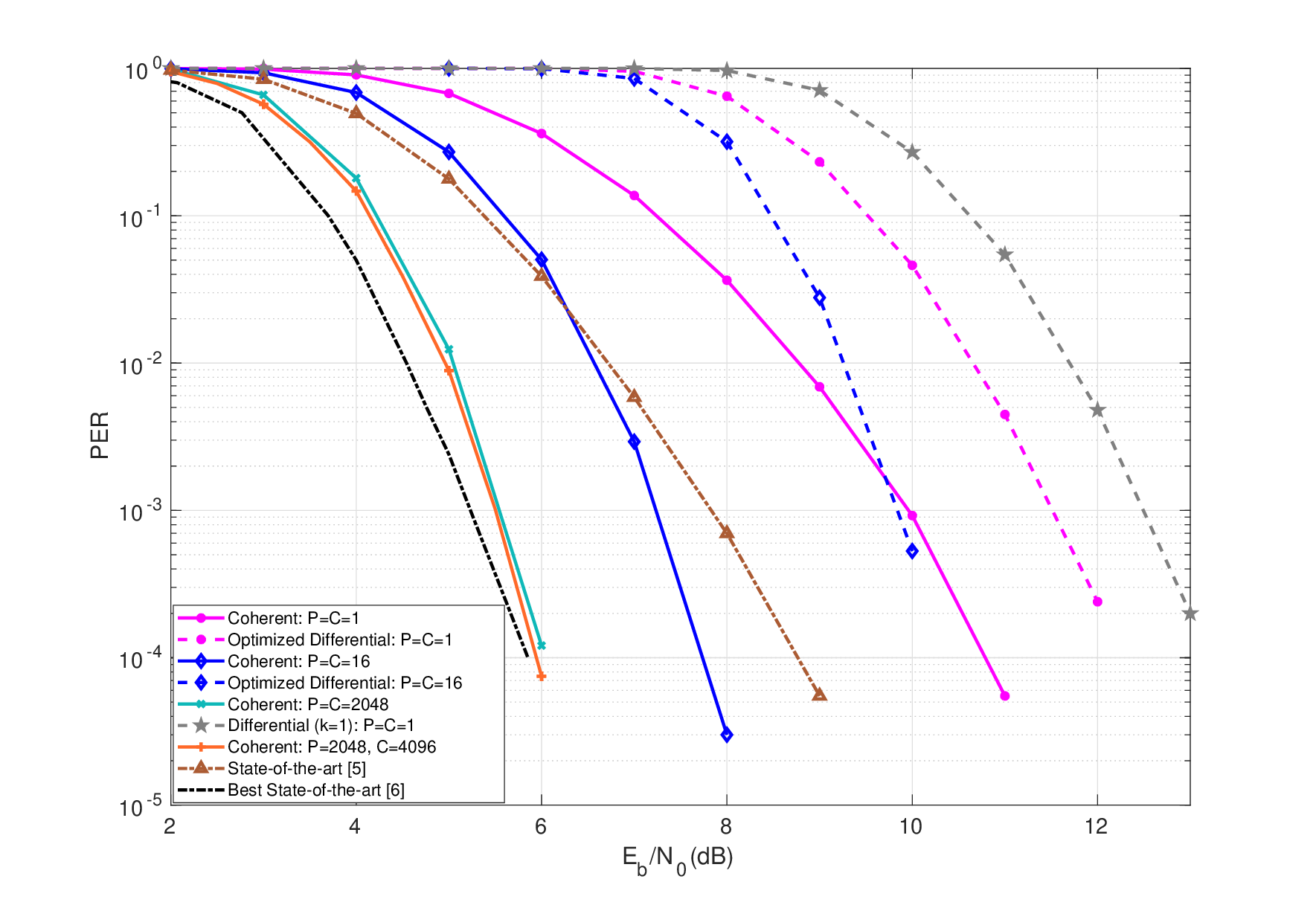}
       
        \caption{Effect of PLVA application with the coherent and optimized differential detectors on the probability of detection in a single-user AWGN channel with varying $(P,C)$ values compared to two state-of-the-art AIS coherent receivers \cite{jointMLSE,threezones2}.}
        \label{fig:noint}
    \end{figure}

    We first illustrate the influence of the list parameters $(P,C)$ on the PLVA detection performance of AIS messages in a single-user AWGN channel. For that we simulated the PLVA with the coherent detector given $P$ equals $\{1,16,2048\}$ and optimized differential detector given $P$ equals $\{1,16\}$ and $C=P$. The packet error rate (PER) is plotted as a function of $E_b/N_0$ in Fig. \ref{fig:noint}. The curve corresponding to the differential detector ($k=1$) with VA ($P=1$) serves as an upper bound on the error rate. We observe that with PLVA having $P=16$ the coherent detector performance improved by approximately 3 dB compared to the VA ($P=1$) at PER$=10^{-3}$. Similarly for the optimized differential detector, with $P=16$, a gain of $\simeq$ 2 dB compared to VA ($P=1$) is obtained. Thus, the data error rate is reduced with the increase of list size for both the coherent and optimized differential detectors. Hence, PLVA shows remarkable improvement of the performance compared to VA in the single-user scenario, especially given the limited computational complexity increase of the LVA with respect to the classical VA approach.
    
    We have also plotted the performance of two other state-of-the-art AIS receivers: the receiver depicted in \cite{jointMLSE} whose complexity is prohibited for practical applications but which achieves the best performance to the best of our knowledge and the receiver described in \cite{threezones2} but assuming perfect synchronization contrary to the work of \cite{threezones2} where synchronization is implemented. We observe that the PLVA with $(P,C)=(16,16)$ outperforms \cite{threezones2} by approximately 1 dB at a PER of $10^{-4}$ and that the PLVA with $(P,C)=(2048,2048)$ approaches the lower bound given by \cite{jointMLSE}, converging beyond 6 dB (PER < $10^{-4}$). In the next section, we show that optimizing $C$ with fixed $P$ further improves the performance in most cases. With $(P,C)=(2048,4096)$, the additional gain is negligible, indicating that the best achievable performance with the PLVA is reached at $P=2048$ and $C\geq2048$.
    
    
    \subsection{Optimization of parameters $(P,C)$ of the PLVA} 
    \label{Detection in the case of the two-users system model}
    
   In order to optimize the parameters $(P,C)$ of PLVA,  we assume having two interfering signals. Focusing on one transmitted signal of interest at a time, and assuming we have only a single interferer with the overlap equal to \{$17\%,83\%$\} of the actual data (data, CRC bits and stuffed bits) approximately corresponding to the worst case scenario of the Class 2 Interference and the best case scenario of the Class 1 Interference considering a representative Sat-AIS scenario. The receiver synchronizes itself with respect to the strongest-received power signal. 
   The PLVA performance depends on the parameters $(P,C)$. Therefore  the optimization of these parameters is a must to meet a trade-off between performance and complexity.
   The optimized $P$ and $C$ values for both the coherent and  differential detectors in a two-user channel model are provided in Table \ref{table:2}, assuming a $-3$ dB power difference, $\{$17\%,83\%$\}$ overlap and neglected Doppler shift difference. The optimized values of $P$ and $C$ remain consistent regardless of whether the overlap is minor or major. For the coherent detector with PLVA, taking $(P,C)=(64,256)$ enables to achieve the same performance gain as  $(P,C)=(256,256)$  with reduction of storage requirement and of processing complexity and time. The same applies for the optimized differential decoder for $(P,C)=(128,512)$ showing a gain similar to that obtained by $(P,C)=(512,512)$.
    
    \begin{table}[bt]
    \caption{Optimized PLVA parameters for AIS signal detection in a Two-user multiple access channel - Target PER of $-3$ dB power difference and $\{$17\%,83\%$\}$ overlap } 
    \centering
    \begin{tabular}{c c c }
    \toprule
     & \textbf{List size} & \textbf{Number of} \\ 
     & $P$ & \textbf{candidates  $C$} \\
     \midrule
      Coherent detector & 64 & 256 \\ 
     Optimized Differential detector &128  &512  \\ 
    \bottomrule
    \end{tabular}
    \label{table:2}
    \end{table}

    \subsection{System-level Simulations}
    \label{Realistic Simulations}

        In this subsection, we consider a more realistic AIS reception scenario from a LEO satellite to provide insights onto the overall reception performance. The satellite altitude is chosen equal to 656.5 km and the AIS transmitting vessels are uniformly distributed in the satellite swath (up to 0 degree ground elevation). For simplicity, the access protocol is modeled under a slotted ALOHA hypothesis (local SOTDMA organization is neglected). This approximation is relevant considering that the satellite swath is much larger than an SOTDMA-organized area and was already considered as a surrogate model for the satellite-AIS reception in \cite{clazzer_analysis_2015}. A number of arrivals per AIS frame is set and the slot choice for each signal is independent and uniformly distributed. The received $C/N_0$, propagation delays and Doppler characteristics are computed assuming static emitters and a simple yet sufficiently representative link budget assuming a pure LOS path (without fading, hence resulting in a limited dispersion of $E_s/N_0$ among users). 

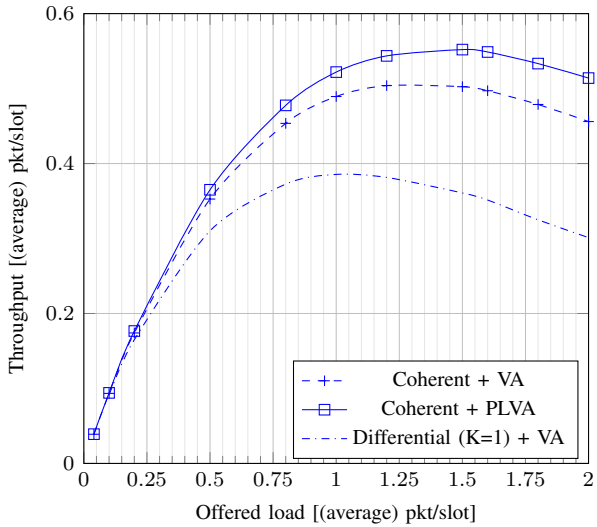
\begin{figure}[t] 
       \centering
       \begin{tikzpicture}
	\pgfplotsset{
        xtick = {0, 0.25, ..., 2},
        xmin=0, xmax=2,
        y axis style/.style={
            yticklabel style=#1,
            ylabel style=#1,
            y axis line style=#1,
            ytick style=#1
       }
    }

    \begin{axis}[
      scale only axis,
    width=0.37\textwidth, 
    height=0.33\textwidth, 
      ymin=0, ymax=0.6,
      ytick = {0, 0.2, ..., 0.6},
      xlabel={Offered load [(average) pkt/slot]}, xtick distance=1, xlabel near ticks,
      ylabel={Throughput [(average) pkt/slot]}, ylabel near ticks, grid=both, minor tick num=4,
      font=\footnotesize, mark options={solid}, legend style={at={(0.98,0.01)},anchor=south east}
    ]

        \addplot[smooth, dashed, mark=+, blue] table[x expr=\thisrowno{0}/2250, y expr=\thisrowno{0}/2250*(1-\thisrowno{1})] {data/per_vs_offered_load_VA.txt};
	    \addlegendentry{Coherent + VA}
		
		\addplot[smooth, mark=square, blue] table[x expr=\thisrowno{0}/2250, y expr=\thisrowno{0}/2250*(1-\thisrowno{1})] {data/per_vs_offered_load_LVA.txt};
		\addlegendentry{Coherent + PLVA} 

        \addplot[smooth, dash dot, blue] table[x expr=\thisrowno{0}/2250, y expr=\thisrowno{0}/2250*(1-\thisrowno{1})] {data/per_vs_offered_load_diffk1.txt};
		\addlegendentry{Differential (K=1) + VA} 

		
	\end{axis}
	
		

        
\end{tikzpicture}
        \caption{Throughput as a function of the offered load, considering Coherent detection using the VA and PLVA with $(P,C)=(64,256)$ and usual Differential detection with $K=1$ using the VA taken as a lower bound.}
        \label{fig:realisticloads}
    \end{figure}

        Considering coherent receivers, the system throughput of the VA and PLVA (with $P,C= (64,256)$) is depicted in Fig.~\ref{fig:realisticloads} with the usual differential detector ($K=1$) curve with VA taken as a lower bound. PLVA significantly outperforms the differential detector used before for AIS. Comparing the two coherent detectors, it can be shown that the PLVA allows an increase in maximum throughput of $\approx 0.05$ (average) pkt/slot and increases the throughput at high system loads, without the use of any interference cancellation strategy. This gain is expected to be even greater with a SIC scheme.

    \vspace{-2 mm}



    \section{Conclusion}
    In this article, we propose the application of the Parallel-List Viterbi algorithm (PLVA) for the purpose of improving the success rate of correct detection of the AIS messages. The performance of PLVA with optimized parameters is studied with the coherent and optimized differential detectors for the decoding of AIS signals. Simulations of the AIS detection in a single user AWGN channel show that the performance of the proposed algorithm is much improved compared to the Viterbi algorithm with the best results exceeding the performance in \cite{threezones2} and approaching the lower bound given by \cite{jointMLSE}, thus showing the effectiveness of PLVA. Finally, simulations under realistic collision conditions with varying channel overload with optimized PLVA is addressed, which leads to promising perspectives concerning the application of PLVA to improve the primary detection step in SIC algorithms.

    \bibliographystyle{IEEEtran}
    \addcontentsline{toc}{section}{Bibliography}
    \bibliography{bib_paper1} 

\end{document}